# Strong coupling in semiconductor hyperbolic metamaterials

*Patrick Sohr, Dongxia Wei, Zhengtianye Wang, and Stephanie Law\**

Department of Materials Science and Engineering, University of Delaware, Newark, Delaware 19702, USA

ABSTRACT Semiconductor-based layered hyperbolic metamaterials (HMMs) house high-wavevector volume plasmon polariton (VPP) modes in the infrared spectral range. VPP modes have successfully been exploited in the weak-coupling regime through the enhanced Purcell effect. In this paper, we experimentally demonstrate strong coupling between the VPP modes in a semiconductor HMM and the intersubband transition of epitaxially-embedded quantum wells. We observe clear anticrossings in the dispersion curves for the zeroth-, first-, second-, and third-order VPP modes, resulting in upper and lower polariton branches for each mode. This demonstration sets the stage for the creation of novel infrared optoelectronic structures combining HMMs with embedded epitaxial emitter or detector structures.

.

**Introduction**

Hyperbolic metamaterials (HMMs) are artificial composite materials composed of subwavelength metal and dielectric features. HMMs have an anisotropic dielectric tensor, where $\varepsilon_{xx}=\varepsilon_{yy}=\varepsilon_\parallel$, $\varepsilon_{zz}=\varepsilon_\perp$ and $\varepsilon_\parallel \cdot \varepsilon_\perp < 0$, which results in an open, hyperbolic isofrequency surface. They are further classified as either a Type I HMM, where $\varepsilon_\parallel>0$ and $\varepsilon_\perp<0$, or a Type II HMM, where $\varepsilon_\parallel<0$ and $\varepsilon_\perp>0$. The point at which an HMM changes from Type I to Type II is referred to as the optical topological transition. Figure 1(a) shows the isofrequency surface for a Type II HMM. The open nature of the isofrequency surface enables HMMs to support the propagation of light with large wavevectors limited only by the subwavelength feature size. In an HMM composed of subwavelength metal and dielectric layers, the modes that support the large wavevector light are called by a variety of names including volume plasmon polaritons (VPPs), hyperbolic volume plasmon polaritons, multilayer plasmons, and bulk plasmon polaritons. In this paper, we will adopt the VPP terminology. VPPs are bulk cavity modes that arise from the coupling of surface plasmon polaritons (SPPs) located at the interfaces of the metal and dielectric layers. The naming convention for the VPP modes in this paper is determined from the number of nodes in their magnetic field profiles (i.e., VPP0 is the mode with zero nodes). These features have made HMMs important for studying a variety of light-matter interactions, including negative refraction[1–5], subwavelength focusing[6–11], modified spontaneous emission[12–17], and slow light[18–20].

HMMs can be used as components of novel optical systems and have already shown significant promise in the weak-coupling regime. This regime is most notably used for increasing the spontaneous emission of emitters through the Purcell effect. The large open isofrequency surface in HMMs gives rise to a large photonic density of states. Emitters coupled to HMMs can therefore have significantly shorter carrier lifetimes or modified emission wavelengths[12,14,16,17]. On the other hand, strong coupling occurs when the coupling strength of two oscillators is greater than their respective decay rates. This regime is characterized by a mode splitting or anticrossing of the dispersion curves of the two oscillators, referred to as the vacuum Rabi splitting[21,22]. In a strongly coupled system away from the coupling point, each mode retains its own character and properties. Near the crossing point, however, the modes become hybridized into an upper branch and a lower branch which can show unusual properties.

Strong coupling was first demonstrated with a single Rydberg atom in a microwave cavity[23], and strong coupling to a single emitter has remained an important area of focus since it opens new possibilities for studying a variety of light-matter interactions, like single atom lasing[24,25] and quantum entanglement[26]. Since the first experimental demonstration, strong coupling has been demonstrated in a variety of systems, primarily focusing on emitters located near metamaterials or within microcavities and photonic crystals[21,27–32]. Strong coupling of quantum dot emitters near an HMM has also been demonstrated[32]. However, strong coupling has yet to be demonstrated between the bulk large wavevector modes—the VPPs—in an HMM and an embedded emitter or absorber. Previous theoretical work demonstrated that it should be possible to strongly couple a VPP mode and the intersubband transition (ISBT) of a quantum well (QW) in a layered semiconductor HMM system[33]. This is particularly useful for the case of infrared HMMs. In the visible spectral range, dye emitters, for example, can easily be coupled to HMMs either by spinning them on top or by adding them to the dielectric layers during the processing of the HMM[12,16,17,34–38]. This is possible because most visible-wavelength HMMs are made of amorphous or polycrystalline materials. In the infrared, however, most emitters and detectors are made from crystalline semiconductor-based materials. The easiest way to couple these devices to HMMs is to grow the entire structure epitaxially.

Here we show experimental evidence of strong coupling between the VPP modes of a Si:InAs/AlSb HMM[39] and the ISBT of an embedded InAs/AlSb QW. A single QW with an ISBT of approximately 14.6μm is embedded within each of the dielectric layers of the HMM. We use metallic gratings to couple the incident light into the VPP modes and map the dispersion of these modes. Our results are compared to a control HMM that has the same structure but without QWs. We observe anticrossing behavior only in the sample with QWs, which conclusively demonstrates that we are observing strong coupling. These results are supported by finite element modeling simulations using Comsol Multiphysics and a calculation of the Rabi splitting energy. This demonstration shows that embedded optoelectronic devices can strongly interact with HMM modes and sets the stage for the creation of semiconductor-based HMMs with embedded infrared detectors, emitters, or other novel optical devices.

**Experimental methods**

The samples discussed in this paper were grown by molecular beam epitaxy (MBE) using a Veeco GenXplor system. The samples were grown on a GaAs substrate using the periodic misfit array growth technique to account for the lattice mismatch between the substrate and the film[40]. A buffer layer of 250nm GaSb/2nm AlSb/250nm GaSb was grown before the HMMs. The buffer layer was grown at 500°C, and the HMM was grown at approximately 425°C as determined by band edge thermometry. The V/III flux ratios were 5/1 for GaAs and GaSb, 10/1 for AlSb, and 20/1 for InAs. Polarization-dependent reflection spectra were measured at an incident angle of 10° using a Bruker Vertex 70V FTIR spectrometer with a Pike Technologies 10Spec accessory. The setup for the FTIR involved a mid-IR source (a SiC glow bar), a KBr beamsplitter, a KRS-5 IR polarizer, and an MCT detector. The polarized reflection measurements were normalized to the polarized 10° reflection measurements of gold. The wavelength range of this setup is 2-25μm. The spectra were then fit using Lorentzian oscillators to obtain the resonant wavelengths of the modes in the HMMs.

**Results and Discussion**

**Design of the structure**

The structure for the Si:InAs/AlSb HMM with embedded InAs QWs is shown in Figure 1(b). The HMMs and the embedded QWs were grown by molecular beam epitaxy (MBE) using silicon-doped InAs as the metallic component[39,41,42]. The Si:InAs/AlSb HMM was designed to have a plasma wavelength in the metal layers of 7.3μm and a total thickness around 2μm, resulting in an optical topological transition point at 9.8μm. The metal and dielectric layers are each 65nm thick with a total of 15 periods (one metal and dielectric layer) in the structure. The HMM is capped with 5nm of undoped InAs. The InAs QWs have a well width of 23nm and an AlSb barrier width of 21nm, resulting in an ISBT from the ground state to the first excited state near 14.6μm. In addition to this sample, a control HMM sample was grown with the same subwavelength structure but without the QWs. Note that this ISBT wavelength is slightly different that that used in modeling, discussed in detail below.

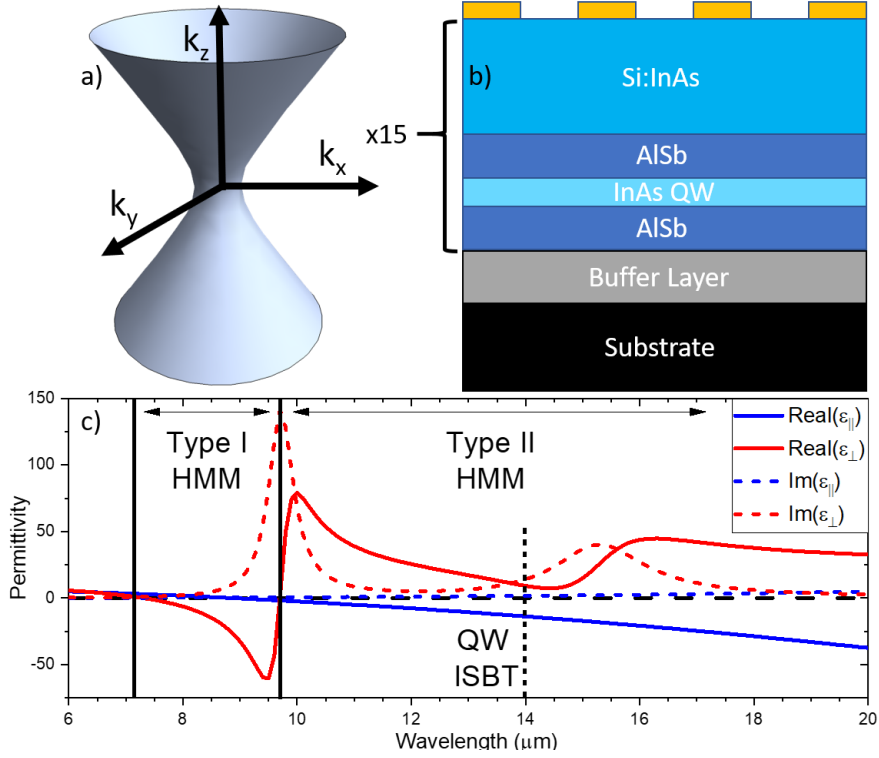

Figure 1: (a) Isofrequency surface for a Type II HMM. (b) Schematic of the structure of the Si:InAs/AlSb HMM with InAs QWs embedded in the dielectric layers. (c) Plot of real and imaginary permittivity versus wavelength for an HMM with a 7.3μm plasma wavelength, an optical topological transition of 9.8μm, and embedded QWs with a 14μm ISBT (to match simulations).

The control HMM and the HMM with QWs were cleaved into multiple pieces and gold grating couplers were applied to the surface using standard electron beam lithography and deposition techniques. Gratings with different periods enable us to map the dispersion curve of the HMMs by varying $k_x$ based on Equation 1:

$$\frac{k_x}{k_0} = \sin\theta + \frac{\lambda}{\Lambda} \qquad \text{(Equation 1)}$$

where $\theta$ is the angle of incidence, and $\Lambda$ is the grating period. The polarization-dependent reflection spectra were measured at an incident angle of 10° using a Bruker Vertex 70v Fourier transform infrared spectrometer. The spectra were then fit using Lorentzian oscillators to obtain the resonant wavelengths of the modes in the HMMs. All the data presented is for transverse magnetic (TM) polarization, since transverse electric (TE) polarization does not observe the anisotropy in HMMs.

**Numerical modeling**

The simulated data was generated using finite element modeling with Comsol Multiphysics. The metal layers were modeled using the Drude model. The dielectric layers with the embedded QWs were modelled as an anisotropic effective medium by adapting the approach laid out by Shekhar and Jacobs[33]. The real and imaginary components of the permittivity of the QW/HMM are shown in Figure 1(c). The QW ISBT was set to 14μm for the simulations. The QW ISBT for the simulation is different from the experimental data because the experimental data had a larger ISBT than initially anticipated. This modifies the location of the strong coupling, but the shape of the curves remains the same for the simulation and experiment.

Figure 2(a) shows the simulated dispersion curves for the control HMM (without QWs). This data is composed of modeled TM-polarized reflection data from HMM samples with different grating periods. Each of the modes is labeled and denoted by a different symbol. The dashed line indicates the position of the QW ISBT in the QW/HMM sample. We can see that the QW ISBT will intersect with VPP0, VPP1, VPP2, and both focusing modes. Focusing modes in an HMM arise due to constructive interference from the grating coupler. The angle at which the light propagates in the HMM is determined by Equation 2[7].

$$\tan(\theta) \cong \sqrt{-\frac{\varepsilon_\parallel}{\varepsilon_\perp}} \qquad \text{(Equation 2)}$$

Light scatters from the edges of the grating and propagates into the HMM at the angle defined by Equation 2. For some specific grating periods, HMM thickness, and wavelength, the light scattered by the edge of one stripe and the light scattered by the edge of the adjacent stripe will impinge upon the HMM/substate interface at the same location. This leads to a larger electric field at this point, which leads to absorption captured in by our model but is not a VPP mode.

In Figure 2(b), the dispersion curve for the HMM with embedded QWs is shown. We see a distinct anti-crossing at the location of the QW ISBT which is indicative of strong coupling between the QW and the VPP modes. Figure 2(b) also provides labels for these new hybridized branches. These were determined from analyzing magnetic field profiles shown in Figure 3 and Figure S1 shown in the Supporting Information. As described previously, when modes strongly couple, they form an upper branch (UB) and a lower branch (LB). In the simulated data, we clearly observe the upper polariton branch for VPP0, VPP1, VPP2, and VPP3 and the lower polariton branch for VPP1, VPP2, VPP3, and VPP4.

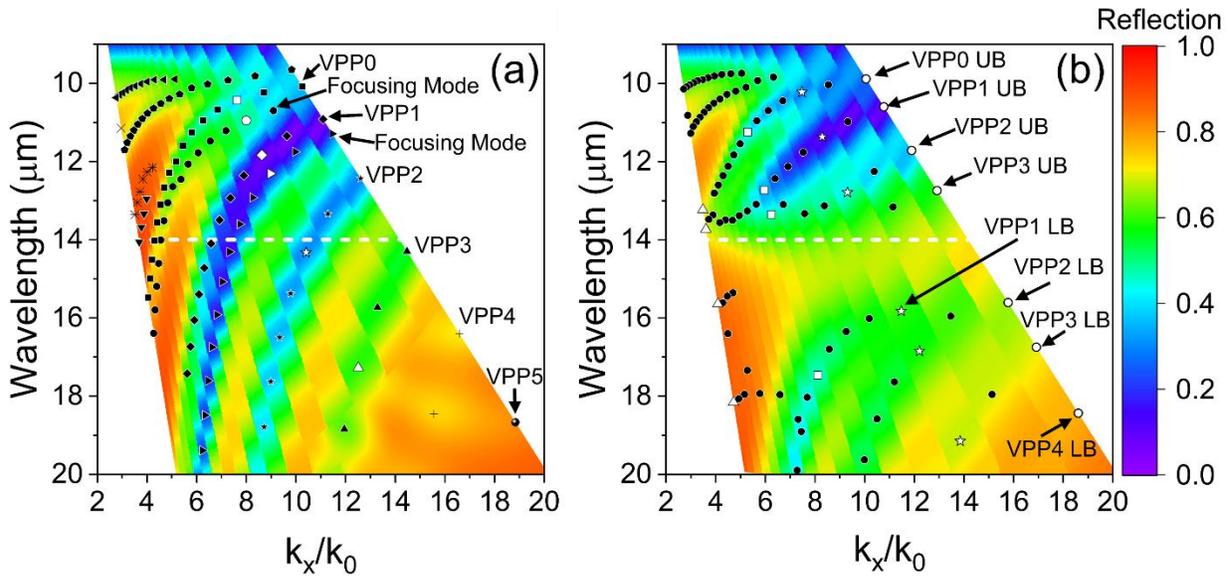

Figure 2: Simulated reflection data. The symbols are the resonant wavelengths for the modes as determined by fitting with Lorentzian oscillators. The white dashed line indicates the ISBT of the InAs QWs. The open symbols indicated which data points have their magnetic field profiles displayed in Figures 3 and 4. (a) shows data for the control HMM with the modes labeled, and

(b) shows data for the HMM with embedded QWs with the upper polariton branches (UB) and lower polariton branches (LB) labeled.

Figure 3(a) shows the magnetic field profiles along the y-axis (out of the page) for the control HMM. The data points corresponding to these mode profiles are indicated by open symbols in Figure 2(a). Figure 3(b) shows the magnetic field profiles along the y-axis for the HMM with QWs embedded in the dielectric layers. The data points are indicated by open stars in Figure 2(b). The grating period for both panels is 1.4μm. The resonant wavelengths for the modes are listed above the profiles.

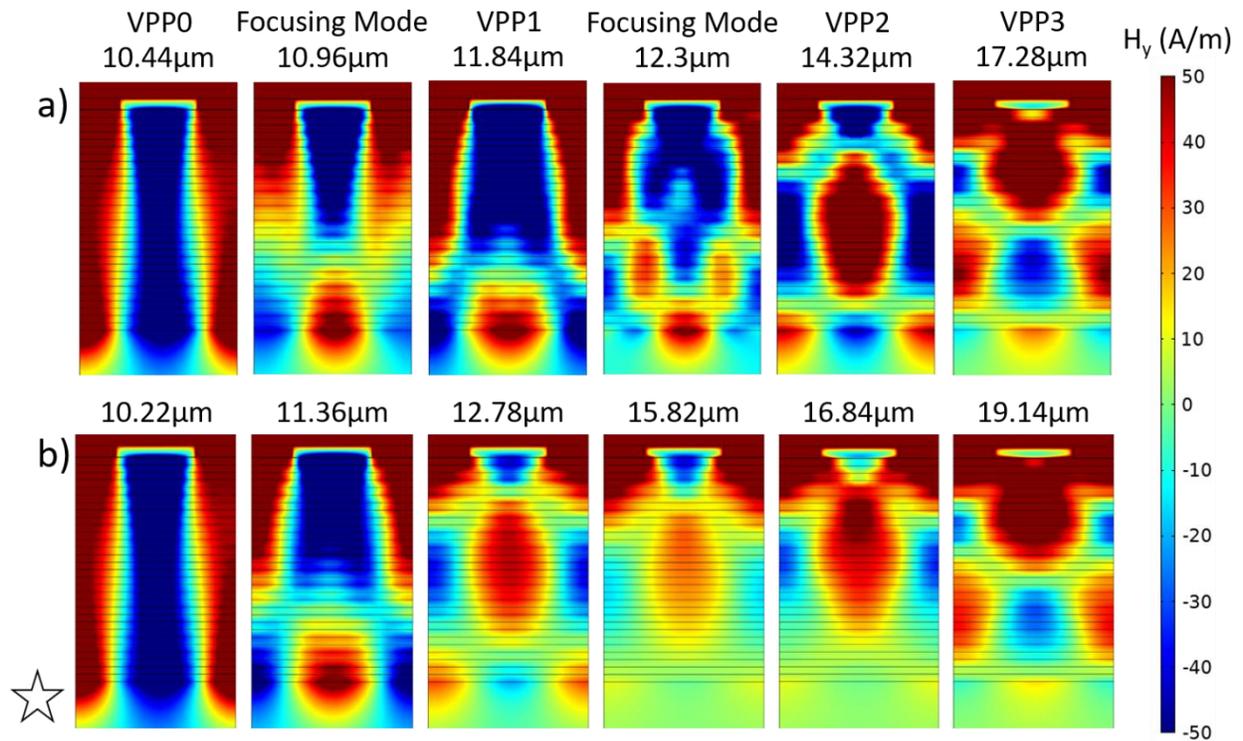

Figure 3: Magnetic field profiles for the modes of the (a) the control HMM with a 1.4μm grating and (b) the HMM with QWs with a 1.4μm grating.

In Figure 3(a), we can see VPP0 through VPP3 with focusing modes interspersed; the modes are labeled above the plots. The magnetic field profiles for the VPP modes align with our expectations[39,43]. For example, the mode labeled as VPP0 has zero nodes in the magnetic field profile. If a line cut is taken through the center of the image, we see that the mode labeled as VPP1 has one node, VPP2 has two nodes, and so on. The VPP modes are standing waves in the bulk of the HMM, so they can be unambiguously identified by the number of nodes in their magnetic field profiles. In Figure 3(b), we identify the modes both by their symmetry and their wavelength. Based on the field profiles and mode position, we assign the first three modes at 10.22mm, 11.36mm, and 12.78mm to the upper hybridized branches caused by coupling between VPP0, VPP1, and VPP2, respectively, with the ISBT. Because the wavelengths of these modes are relatively far removed from the wavelength of the QW absorption, they are not strongly hybridized and retain most of the VPP character though their absolute positions are blueshifted from the unhybridized modes. The blueshift is expected for the upper branches of a hybridized system.

A similar argument can be made for the last two modes at 16.84mm and 19.14mm. Based on their field profiles and mode positions, they are likely the lower hybridized modes of VPP2 and VPP3, respectively. These modes are slightly redshifted, again as expected for the lower branches in a hybridized system. The fourth mode—at 15.82μm—is likely the lower hybridized mode of VPP1 due to the position of the mode. The field profile looks distinctly different from the unperturbed VPP1 modes because this mode is close enough to the strong coupling point that it has begun to take on strong characteristics of the QW ISBT. Magnetic field profiles for additional hybridized modes are shown in the Supporting Information.

We do not clearly observe the focusing modes in the HMM/QW system. This can be explained by noting that the location of the focusing modes is strongly dependent on the real part of the $\varepsilon_\parallel$ and $\varepsilon_\perp$ of the HMM, as noted in Equation 2. Figure 1(c) shows that embedding a QW into the HMM drastically alters the real part of $\varepsilon_\perp$, which can alter the resonance of the focusing modes out of the field of view. Additionally, the focusing modes are relatively weak and narrow compared to the VPP modes, which could make them difficult to observe in the strongly coupled system.

**Experimental data**

Figure 4(a) shows the experimental dispersion curve for the control HMM (without QWs). This data is composed of TM-polarized reflection data from HMM samples with different grating periods. Each of the modes is labeled and denoted by a different symbol. The dashed white line shows where the ISBT of the InAs QWs is. Note that the view into wavevector space is smaller than what was shown for the simulated data. In addition, the QW ISBT is at 14.6μm instead of the simulated 14μm. Figure 2(b) shows the experimental dispersion curve for the HMM with QWs embedded in the dielectric layers. We observe a clear anticrossing at the absorption wavelength corresponding to the ISBT of the QWs just like in the simulated data, indicating strong coupling between the QW ISBT and the HMM VPP modes. The modes are all denoted with the same symbol in Figure 4(b). This is intentional, since the significant hybridization between the QW and the VPP and focusing modes makes it challenging to conclusively to assign specific points to specific hybridized branches without further analysis.

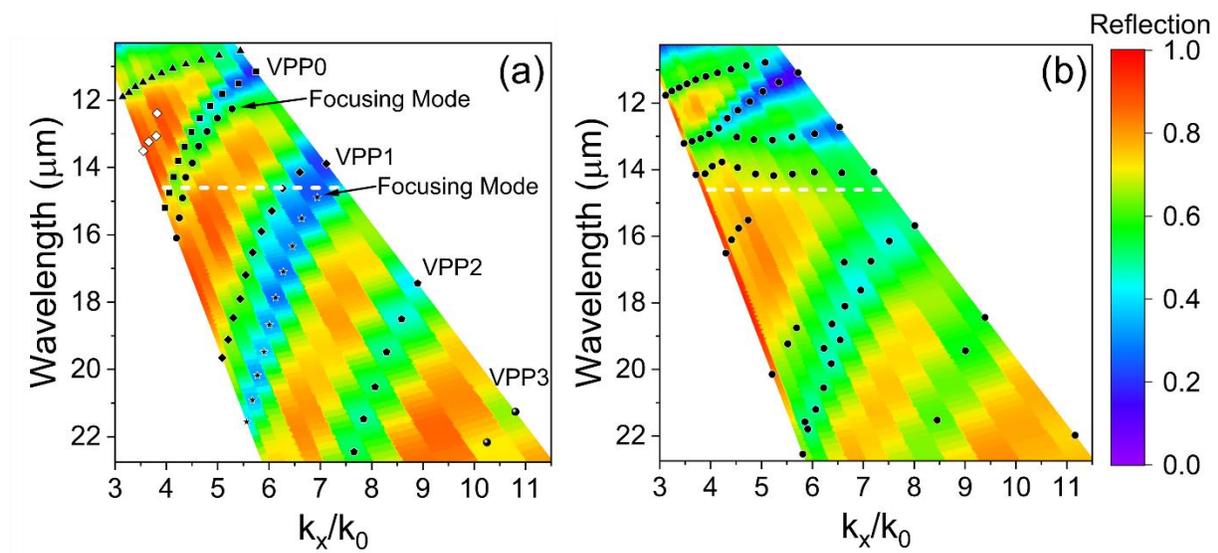

Figure 4: Experimental reflection data. The symbols are the resonant wavelengths for the modes as determined by fitting the reflection data with Lorentzian oscillators. The white dashed line indicates the ISBT of the InAs QWs. (a) shows data for the control HMM with the modes labeled, and (b) shows data for the HMM with embedded QWs.

In order to conclusively demonstrate that we observe strong coupling in our system, we performed a Rabi splitting calculation. Conceptually, we can model the system as a set of coupled oscillators where the Rabi splitting energy determines the strength of the coupling. If the Rabi splitting energy is larger than the losses in the system, the system is strongly coupled. To perform this calculation, we followed the methods of [33]. The energies of the upper and lower polariton branches ($E_{U,L}$) are given by

$$E_{U,L}(k) = \frac{1}{2}\left[E_{VPP}(k) + E_{ISBT} \pm \sqrt{\hbar\Omega + (E_{VPP}(k) - E_{ISBT})^2}\right] \quad \text{(Equation 3)}$$

where $E_{VPP}(k)$ is the wavevector-dependent energy of each VPP mode, $E_{ISBT}$ is the energy of the quantum well intersubband transition, and $\hbar\Omega$ is the Rabi splitting energy. To extract the Rabi splitting energy for each VPP mode, we first model the VPP mode dispersions for the uncoupled system (Figures 1(a) and 5(a)). We used a phenomenological fitting equation

$$E_{VPP}(k) = a - bc^k \quad \text{(Equation 4)}$$

where $a$, $b$, and $c$ are all fitting parameters. This phenomenological model produced a good fit ($R^2 > 0.99$) for all uncoupled VPP mode dispersions for both the simulated and experimental data. Using that model for $E_{VPP}(k)$ and the known QW ISBT energy, we then used Equation 3 to fit the coupled dispersion curves for both the simulated and experimental data, as shown in Figure 5(a) and (b), respectively.

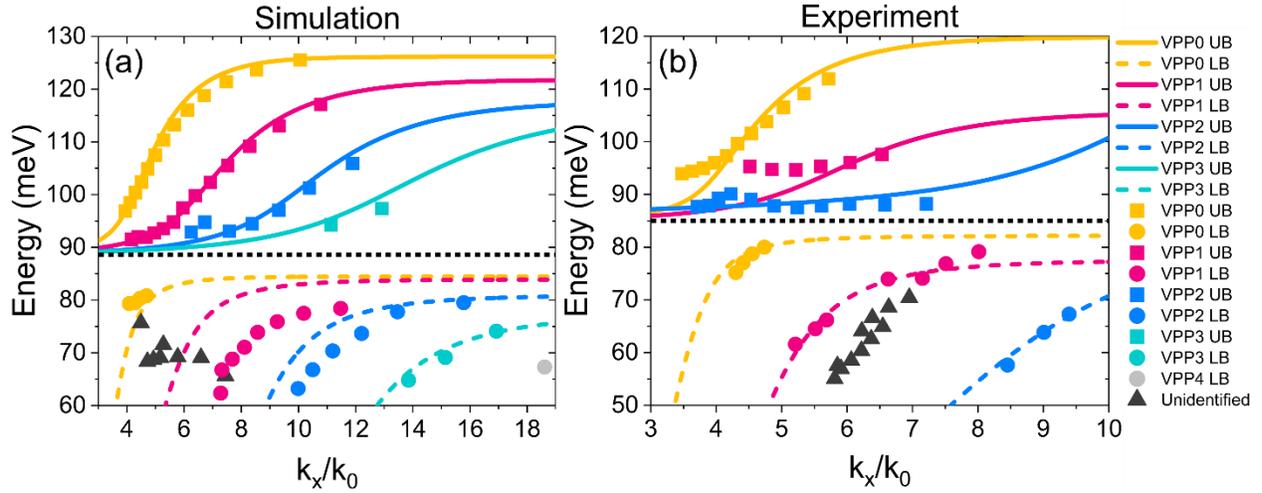

Figure 5: Rabi splitting calculation for the (a) simulated and (b) experimental spectra. Solid lines are the calculated upper polariton branches, dashed lines are the calculated lower polariton branches, squares are the simulated or experimental upper polariton branch data points, circles are the simulated or experimental lower polariton branch data points, and triangles are unidentified resonances. The data points presented here are the same ones presented in Figure 2(b) and Figure 5(b). The black dotted line is the QW ISBT, which falls at 14μm=88.6meV in the simulation and 14.6μm=85meV in the experiment.

Overall, we observe good agreement between the dispersion curves calculated using Equation 3 (solid lines for the upper polariton branches; dashed lines for the lower polariton branches), and the data points (squares for the upper polariton branches; circles for the lower polariton branches). However, for both the simulated data and the experimental data, we get better agreement for the upper branches than for the lower. We observe an especially large discrepancy for the lower polariton branch of VPP1. This may be caused by the presence of the unidentified modes, denoted with gray triangles. These appear in both the simulated and experimental data and may be related to the focusing modes described earlier. Despite this, we generally have a good match between calculation and simulated or experimental data points.

The extracted Rabi splitting energies for the simulation data are 25meV, 25meV, 30meV, and 35meV for VPP0, VPP1, VPP2, and VPP3, respectively. For the experimental data, the Rabi splitting energies are 20meV, 25meV, and 30meV for VPP0, VPP1, and VPP2, respectively. The electron scattering losses for the QW are approximately 7meV, and the scattering losses for the HMM are also approximately 7meV. Together, the loss in the system is 14meV, which is less than the Rabi splitting energies, indicating that the material is in the strong coupling regime for all VPP modes considered.

## **Conclusions**

We have demonstrated strong coupling between the bulk large wavevector volume plasmon polariton modes in a layered, semiconductor HMM and the intersubband transition of epitaxially embedded QWs. We observe the upper and lower hybridized branches associated with strong coupling between three different VPP modes and the embedded QW. These results are supported by a Rabi splitting calculation, which indicates that the coupling energy between the QW and the VPP modes is 20-35meV, depending on which VPP mode is being considered. This demonstration clearly shows that VPP modes will interact with and strongly couple to epitaxially-embedded structures. This is particularly important for scientists and engineers working in the infrared spectral range, where most optoelectronic devices rely on the use of semiconductor-based epitaxial device stacks. Device technologies can now be coupled to and enhanced by interaction with hyperbolic metamaterial modes, and the entire structure can be grown in a single molecular beam epitaxy deposition, potentially leading to more efficient infrared optoelectronic devices.

AUTHOR INFORMATION

**Corresponding Author**

*slaw@udel.edu

**Author Contributions**

S.L. conceived and supervised the project. P.S. and D.W. grew, fabricated, and measured the samples and performed numerical modeling. Z.W. assisted with sample fabrication. All authors discussed the results. S.L. and P.S. wrote the manuscript with comments from all authors.

**Funding Sources**

This research was supported by the U. S. National Science Foundation (Grant No. 1606673). Z. W. acknowledges funding from the U.S. Department of Energy (DOE), Office of Science, Office of Basic Energy Sciences under Award No. DE-SC0017801. The authors acknowledge the use of the Materials Growth Facility (MGF) at the University of Delaware, which is partially supported by the National Science Foundation Major Research Instrumentation under Grant No. 1828141 and is a member of the Delaware Institute for Materials Research (DIMR).

**Notes**

The authors declare no competing financial interest

ACKNOWLEDGMENT

This research was supported by the U. S. National Science Foundation (Grant No. 1606673). Z. W. acknowledges funding from the U.S. Department of Energy (DOE), Office of Science, Office of Basic Energy Sciences under Award No. DE-SC0017801. The authors acknowledge the use of the Materials Growth Facility (MGF) at the University of Delaware, which is partially supported by the National Science Foundation Major Research Instrumentation under Grant No. 1828141 and is a member of the Delaware Institute for Materials Research (DIMR).

SUPPORTING INFORMATION

Supporting Information: magnetic field profiles for additional simulated samples.

TOC Graphic

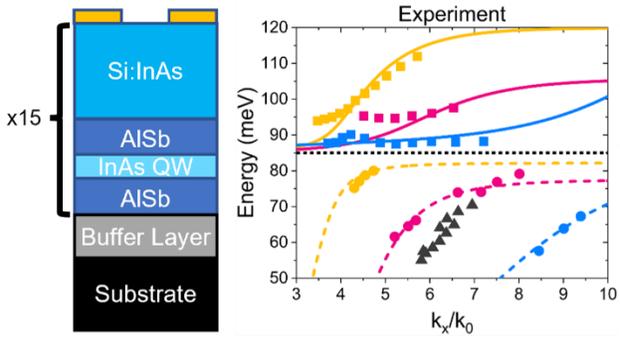

# Supporting information: Strong coupling in semiconductor hyperbolic metamaterials

*Patrick Sohr, Dongxia Wei, Zhengtianye Wang, and Stephanie Law\**

Department of Materials Science and Engineering, University of Delaware, Newark, Delaware 19702, USA

*slaw@udel.edu

**Magnetic field profiles for hybridized modes**

To map the evolution of the hybridized modes as they move closer in wavevector to the QW absorption, Figure S1 shows the magnetic field profiles along the y-axis for the HMM with QWs embedded in the dielectric layers for four different grating periods: 1μm, 1.4μm, 2.2μm, and 4.0μm. The data points are indicated by open symbols (shown next to the magnetic field profiles) in Figure 2(b) in the main text. Figure S1(b) is the same as Figure 3(b) in the main text and is reproduced here for ease of comparison. In Figure S1(a), we can see that all the field profiles for the first four modes and the last mode have the same symmetry as the unhybridized VPP modes, which makes their identification straightforward. The two remaining modes at 15.60μm and 16.74μm can be identified as the lower polariton branches of VPP2 and VPP3, respectively, by their position and symmetry. As the grating period is increased, the modes take on more features of the QW ISBT, which make them challenging to differentiate. In Figure S1(c), the upper hybridized modes are easily discernible, while the lower hybridized modes are not. In Figure S1(d), all the modes have primarily QW ISBT characteristics and look very similar.

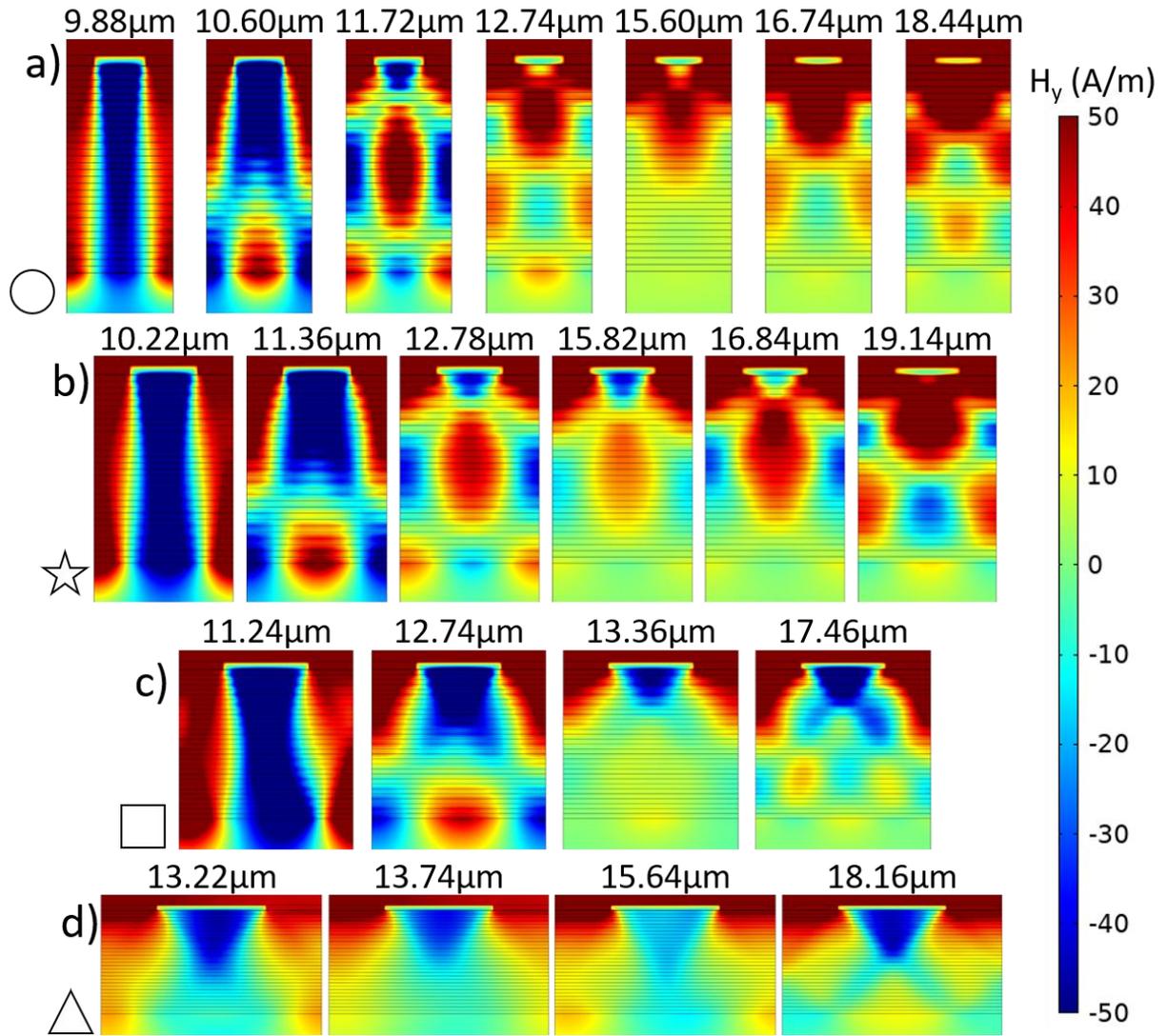

Figure S1: Magnetic field profiles for the modes of the HMM with QWs with a) 1μm grating, b) 1.4μm grating, c) 2.2μm grating, and d) 4μm grating.